\documentclass{article}

\usepackage{mathptmx}       
\usepackage{helvet}         
\usepackage{courier}        
\usepackage{makeidx}         
\usepackage{graphicx}        
\usepackage{multicol}        

\makeindex             

\title{Bekenstein and the Holographic Principle: Upper bounds for Entropy
}
\author{Antonio Alfonso-Faus\thanks{E-mail: aalfonsofaus@yahoo.es} \\ 
Escuela de Ingenier\'{\i}a Aeron\'autica y del Espacio \\
Plaza del Cardenal Cisneros, 3, 28040 Madrid, Spain\\
\and M\`arius Josep Fullana i Alfonso\thanks{E-mail: mfullana@mat.upv.es} \\
Institut de Matem\`atica Multidisciplin\`aria, \\
Universitat Polit\`ecnica de Val\`encia \\
Cam\'{\i} de Vera s/n, 46022 Val\`encia, Spain}

\begin{document}

\maketitle

\abstract{Using the Bekenstein upper bound for the ratio of the entropy $S$ of any bounded system, with energy $E = Mc^2$ and effective size $R$, to its energy $E$ i.e. $S/E < 2\pi k R/\hbar c$, we combine it with the holographic principle (HP) bound ('t Hooft and Susskind) which is $S \le \pi k c^3R^2/\hbar G$. We find that, if both bounds are identical, such bounded system is a black hole (BH). For a system that is not a BH the two upper bounds are different. The entropy of the system must obey the lowest bound. If the bounds are proportional, the result is the proportionality between the mass M of the system and its effective size $R$. When the constant of proportionality is $2G/c^2$ the system in question is a BH, and the two bounds are identical. We analyze the case for a universe. Then the universe is a BH in the sense that its mass $M$ and its Hubble size $R \approx ct$, t the age of the universe, follow the Schwarzschild relation $2GM/c^2 = R$. Finally, for a BH, the Hawking and Unruh temperatures are the same. Applying this to a universe they define the quantum of mass 
$\sim 10^{-66} g$ for our universe.}

\section{Introduction}
\label{sec:1}

In 1981 Bekenstein \cite{1} found an upper bound for the ratio of the entropy $S_B$ to the energy $E = Mc^2$ of any bounded system with effective size $R$:
\begin{equation}
S_B /E < 2 \pi k R / \hbar c
\label{eq1}
\end{equation}

About ten years later \cite{2,3} a HP was proposed giving a bound for the entropy $S_h$ of a bounded system of effective size $R$ as

\begin{equation}
S_h \le \pi k c^3R^2/ \hbar G
\label{eq2}
\end{equation}

The Bekenstein bound (\ref{eq1}) is proportional to the product $MR$, while the HP bound (\ref{eq2}) is proportional to the area $R^2$. If the two bounds are identical (hence $M$ is proportional to $R$) here we prove that the system obeys the Schwarzschild condition for a BH. We analyze this conclusion for the case of a universe with finite mass $M$ and a Hubble size $R \approx ct$, $t$ being the age of the universe. $M$ and $R$ are obviously the maximum values they can have in our universe.
Also, for the case of a BH we find that the Hawking and Unruh temperatures are the same. Then, for our universe we obtain the mass of the gravity quanta, of the order of $10^{-66} g$.

\section{Consequences of the identification of the two bounds}
\label{sec:2}
Identifying the bounds (\ref{eq1}) and (\ref{eq2}) we get
\begin{equation}
2M = c^2R/G
\label{eq3}
\end{equation}

\noindent
which is the condition for the system $(M, R)$ to be a BH. Then, its entropy is given by the Hawking relation \cite{4} 
\begin{equation}
S_H = 4 \pi k / \hbar c G M^2
\label{eq4}
\end{equation}
that coincides with the two bounds (\ref{eq1}) and (\ref{eq2}).     
The mass of the universe $M_u$ is a maximum. And so is its size $R$. A bounded system implies a finite value for both. Using present values for 
$M_u \approx 10^{56}g$ and $R \approx  10^{28}cm$ they fulfill the Schwarzschild condition (\ref{eq3}). 
This is an evidence for the Universe to be a BH \cite{5}. And its entropy today is about $10^{122} k$.
\section{The case for the Unruh and Hawking temperatures}
\label{sec:3}

The fact that a BH has a temperature, and therefore an entropy, Hawking \cite{4} , implies that an observer at its surface, or event horizon, sees a perfect blackbody radiation, a thermal radiation with temperature $T_H$ given by
\begin{equation}
T_H = \hbar c^3 / (8 \pi GMk)
\label{eq5}
\end{equation}

\noindent 
where $M$ is the mass of the BH. This observer feels a surface gravitational acceleration $R''$. Following Unruh effect \cite{6} an accelerated observer also sees a thermal radiation at a temperature $T_U$, proportional to the acceleration $R''$ and given by
\begin{equation}
T_U = \hbar R''  / (2 \pi ck)
\label{eq6}
\end{equation}

Based upon the similarity between the mechanical and thermodynamical properties of both effects, (\ref{eq5}) and (\ref{eq6}), we identify both temperatures and find the relation:

\begin{equation}
R'' = c^4 / (4GM)
\label{eq7}
\end{equation}

Identifying the Unruh acceleration to the surface gravitational acceleration: 

\begin{equation}
R'' = GM / R^2
\label{eq8}
\end{equation}
and substituting in (\ref{eq7}) we finally get
\begin{equation}
2GM / c^2 = R
\label{eq9}
\end{equation}
This is the condition for a BH. Since the Hawking temperature refers to a BH this result confirms the validity of the identification of the two temperatures, (\ref{eq5}) and (\ref{eq6}), as well as the interpretation of the Unruh acceleration in (\ref{eq8}).

\section{Application of the cosmological principle (CP)}
\label{sec:4}

The CP may be stated with the 2 special conditions of the universe: it is homogeneous and isotropic. This means that, on the average, all places in universe are equivalent (at the same {\it time}) and that observing it at one location it looks the same in any direction. The CMBR is a good example, a blackbody radiation at about $2.7 K$. This implies there is no center of the universe, or equivalently, any local place is a center. 
We have seen the universe may be taken as a BH, and so it makes sense to think there must be an event horizon, a 2 dimensional bounding surface around each observer. If all places in universe are equivalent then all places can say this, and the natural event horizon is the Hubble sphere, with radius $R$ about $c / H \approx 10^{28} cm$ today. Then at any point it can be interpreted as a 2 dimensional spherical surface, maybe in a {\it virtual} sense. Following the HP, all the information of the 3 dimensional world, as we see it, is contained in this spherical surface. And any observer, following the CP, can be seen as being at the center of the 3 dimensional {\it sphere}. To combine both principles, the cosmological and the holographic, we can think of an isotropic, spherically symmetric, acceleration present at each point in the universe given by (\ref{eq8}). This is a change of view from a 3 dimensional world to a 2 dimensional one. Also we have an isotropic temperature given by
\begin{equation}
T = \hbar c^3 / (8 \pi GMk) = (1/4 \pi k) 1/R \approx 10^{-29} K  
\label{eq10}
\end{equation}

This is the temperature of the gravitational quanta \cite{7} at present time. The equivalent mass of one quantum of gravitational potential energy is then from (\ref{eq10}) found to be about $10^{-66} g$. This may be interpreted as the ultimate quantum of mass. Its wavelength (in the Compton sense) is of the order of the size of the universe, $ct \approx 10^{28} cm$. The scale factor between the Planck scale and our universe today is about $10^{61}$. Multiplying the temperature found in (\ref{eq10}) by this numerical scale factor we get the Planck's temperature $T_p \approx 10^{32} K$ at the Planck's time $10^{-44} s$, when the universe had the Planck's size $l_p \approx 10^{-33} cm$.

\section{Conclusions}

The universe can be seen as a BH. We can interpret each observer as beeing at the center of a sphere, with the Hubble radius. From this we interpret the spherical surface as the event horizon, a 2 dimensional surface that follows the physics of the HP.
The isotropic acceleration present at each point in the universe, and given by (\ref{eq8}), implies no distortion for the spherically distributed acceleration, as imposed by the CP. However, the presence of a nearby important mass, like the sun, will distort this spherically symmetric picture. With respect to the probes Pioneer 10/11, that detected an anomalous extra acceleration towards the sun of value \cite{8}
$(8.74 \pm 1.33) \times 10^{-8} cm/s^2$, we see this value is only a bit higher than that predicted by (\ref{eq8}): 
$7.7 \times 10^{-8}cm/s^2$. This difference is an effect that can be explained by the presence of the sun converting the isotropic acceleration to an anisotropic one.
Similarly there may be a factor, due to the influence of nearby masses (i.e. a massive BH at the center of the galaxy), in the cases of the observed rotation curves in spiral galaxies. They imply that the speed of stars, instead of decreasing with distance $r$ from the galactic center, is constant or even increases slowly when far from the central luminous object \cite{9}. For globular clusters \cite{10}, where no dark matter is expected, we have stronger evidence in support of the existence of the acceleration field. Also the escape velocity at the Sun location, with respect to our galaxy, is higher than expected. The earth-moon distance increases with time and there is a residual part not explained by tidal effects. And the same occurs for the planets in the solar system. We present this evidence in support of the universal field of acceleration \cite{11}, $R"$.
Finally, for a BH we find Hawking and Unruh temperatures equal. For our universe we obtain the mass of the gravity quanta, $\sim 10^{-66} g$, with a wavelength corresponding to its size. It may be identified with the bit, with entropy $k$.


\begin{thebibliography}{99.}
\bibitem{5} Alfonso-Faus, A., Astrophys. and Space Sci. {\bf 325}, 113-117 (2010).
\bibitem{11} Alfonso-Faus, A.: arXiv:0708.0308 (2010).
\bibitem{7} Alfonso-Faus, A.:  arXiv: 1105.3143 (2011).
\bibitem{8} Anderson, J. D., et al., Phys. Rev. Lett. {\bf 81}  2858-2861 (1998).
\bibitem{1} Bekenstein, J.D., Phys. Rev. D {\bf 23} (2), 287-298 (1981). 
\bibitem{9} Drees, M., Chung-Lin, S.,  JCAP {\bf 0706}, 011 (2007).
\bibitem{4} Hawking, S.W., Nature {\bf 248}, 30 (1974).
\bibitem{10} Scarpa, R., Falomo, R.,  arXiv:1006.4577 (2010).
\bibitem{3} Susskind, L., J. Math. Phys. {\bf 36}, 6377 (1995).
\bibitem{2} 't Hooft, G., {\it Dimensional Reduction in Quantum Gravity}. arXiv:gr-qc/9310026 (1993).
\bibitem{6} Unruh, W.G., Physical Review D {\bf 14} (4), 870 (1976).
\end{thebibliography}
\end{document}